\documentclass[10pt,letterpaper]{article}
\usepackage[top=0.85in,left=2.75in,footskip=0.75in,marginparwidth=2in]{geometry}

\usepackage[utf8x]{inputenc}
\usepackage{tipa}
\usepackage{newunicodechar}
\usepackage{indentfirst}
\usepackage{enumitem}
\usepackage{caption}
\usepackage{subcaption}
\usepackage{tabularx}

\usepackage{cite}
\usepackage{hyperref}
\hypersetup{
    colorlinks=true,
    linkcolor=blue,
    filecolor=magenta,      
    urlcolor=cyan,
    pdftitle={Overleaf Example},
    pdfpagemode=FullScreen,
    }
    
\usepackage{nameref,hyperref}

\usepackage[right]{lineno}

\usepackage{microtype}
\DisableLigatures[f]{encoding = *, family = * }

\raggedright
\usepackage{changepage}

\usepackage[aboveskip=1pt,labelfont=bf,labelsep=period,singlelinecheck=off]{caption}

\makeatletter
\renewcommand{\@biblabel}[1]{\quad#1.}
\makeatother

\usepackage{lastpage,fancyhdr,graphicx}
\usepackage{epstopdf}

\fancyheadoffset[L]{2.25in}
\fancyfootoffset[L]{2.25in}

\usepackage{color}

\definecolor{Gray}{gray}{.25}

\usepackage{graphicx}

\usepackage{sidecap}

\usepackage{wrapfig}
\usepackage[pscoord]{eso-pic}
\usepackage[fulladjust]{marginnote}
\reversemarginpar

\begin{document}
\vspace*{0.35in}

\begin{flushleft}
{\Large
\textbf\newline{Comparison of algorithms used in single-cell transcriptomic data analysis}
}
\newline
\\
Jafar Isbarov\textsuperscript{1}, Elmir Mahammadov\textsuperscript{2,}\textsuperscript{3,}\textsuperscript{4}
\\
\bigskip
\bf{\textsuperscript{1}} Azerbaijan State Oil and Industry University, Baku, Azerbaijan
\\
\bf{\textsuperscript{2}} Institute of Computational Biology, Helmholtz Zentrum München, Neuherberg, Germany
\\
\bf{\textsuperscript{3}} Institute of Epigenetics and Stem Cells, Helmholtz Zentrum München, Munich, Germany
\\
\bf{\textsuperscript{4}} Institute of Functional Epigenetics, Helmholtz Zentrum München, Neuherberg, Germany

\end{flushleft}

\section*{Abstract}
Single-cell analysis is an increasingly relevant approach in "omics'' studies. In the last decade, it has been applied to various fields, including cancer biology, neuroscience, and, especially, developmental biology. This rise in popularity has been accompanied with creation of modern software, development of new pipelines and design of new algorithms. Many established algorithms have also been applied with varying levels of effectiveness. Currently, there is an abundance of algorithms for all steps of the general workflow. While some scientists use ready-made pipelines (such as Seurat), manual analysis is popular, too, as it allows more flexibility. Scientists who perform their own analysis face multiple options when it comes to the choice of algorithms. We have used two different datasets to test some of the most widely-used algorithms. In this paper, we are going to report the main differences between them, suggest a minimal number of algorithms for each step, and explain our suggestions. In certain stages, it is impossible to make a clear choice without further context. In these cases, we are going to explore the major possibilities, and make suggestions for each one of them.

\section*{Introduction}

\paragraph{}
Quantitative analysis of cell contents - mostly DNA, RNA and protein - has been a major part of molecular biology since the emergence of the field. Studying the protein content of the cell is helpful when trying to uncover dynamics of its metabolism. Observing changes in DNA sequence through the generations allows us to understand the evolutionary processes at a cellular level. Comparing the amount of different RNA molecules enables us to fill the gap between these two separate pieces of information by unveiling the expression patterns of the cell. \\
Our report concentrates on single-cell RNA sequencing analysis - a revolutionary combination of molecular and computational techniques that opens up new dimensions in our understanding of tissues and cell types via more detailed analyses of spatial and temporal variations in gene expression. Before we get started, it is important to understand both conceptual and technical differences between single-cell analysis and its traditional counterpart.

\subsection*{The tradition: bulk analysis}
Molecular analyses rely on sample abundance. In order to sequence a single gene, we need multiple copies. The same thing applies to RNA sequences, too. Amplification techniques such as PCR allows us to sequence small DNA and RNA samples. We can reproduce and sequence particular nucleic acids, but we can not reproduce individual cells. Instead, a tissue has to be assumed to be homogeneous and analyzed in bulk. The power of this approach lies in its simplicity - so does its weakness. Bulk analysis did, and still does, provide insightful information about expression levels of genes in a tissue. But this strips us of our ability to analyze cells individually. This limitation is not merely principal. It comes with a set of problems in our understanding of cell states. Simpson's Paradox is an interesting example of this. It refers to situations in statistical analysis where a trend is observed in separate groups of data, but reverses or disappears altogether when the said groups are combined. For example, expression levels of two genes that otherwise look negatively correlated may turn out to have positive correlation after proper clustering \cite{Trapnell2015}.
\\
Only in the 21\textsuperscript{st} century did we begin to approach these problems. Progress in molecular techniques, coupled with advances in computer science and information technologies set the stage for a new generation of analysis tools to be developed.

\subsection*{The novelty: single-cell analysis}
As a result of progress in amplification techniques and cell capture/isolation instruments, we are now able to perform RNA sequencing on individual cells. RNA-seq is not the only relevant method of measuring cell state; others include Hi-C (method for measuring chromatin conformation),  which is useful for determining the physical interaction of genome regions \cite{Hakim2012}; bisulfite sequencing, and DNase I hypersensitivity sequencing \cite{Trapnell2015}. Each one of these can now be performed on a single-cell scale. However, single-cell RNA sequencing analysis has been our sole concentration through the program, as others are more expensive and can have lower sensitivity. Before reporting our research and findings, we first provide a brief introduction to the field.

\subsection*{Single-cell sequencing}
Transcriptome of the cell has to be sequenced before we can begin computational analysis. Single-cell sequencing is composed of the following steps: (i) cell isolation; (ii) cell lysis; (iii) reverse transcription; (iv) amplification; (v) transcript coverage; (vi) strand specificity; and (vii) Unique Molecular Identifiers (Chen et al., 2019). There are several platforms that enable RNA (and cDNA) sequencing at single-cell resolution. They can be divided into two groups: (1) molecular tag-based methods and (2) full-length methods. 10x Genomics Chromium system and MARS-seq belong to the first group. They are both cheaper, faster, and allow sequencing of more cells at the same time, but provide limited coverage of the transcriptome. The second group, which includes SMART-seq and SMART-seq2, covers the entire transcriptome, but these methods come with immense time and money costs. Additionally, they allow sequencing of a lower number of cells at the same time. Both types of platforms have their place in single-cell studies, however, due to their lower cost and higher speed, molecular tag-based methods are used more widely \cite{See2018}.

\subsection*{Stages of single-cell RNA sequencing (scRNA-seq) analysis}
\begin{figure}
    \centering
    \includegraphics[width=\textwidth]{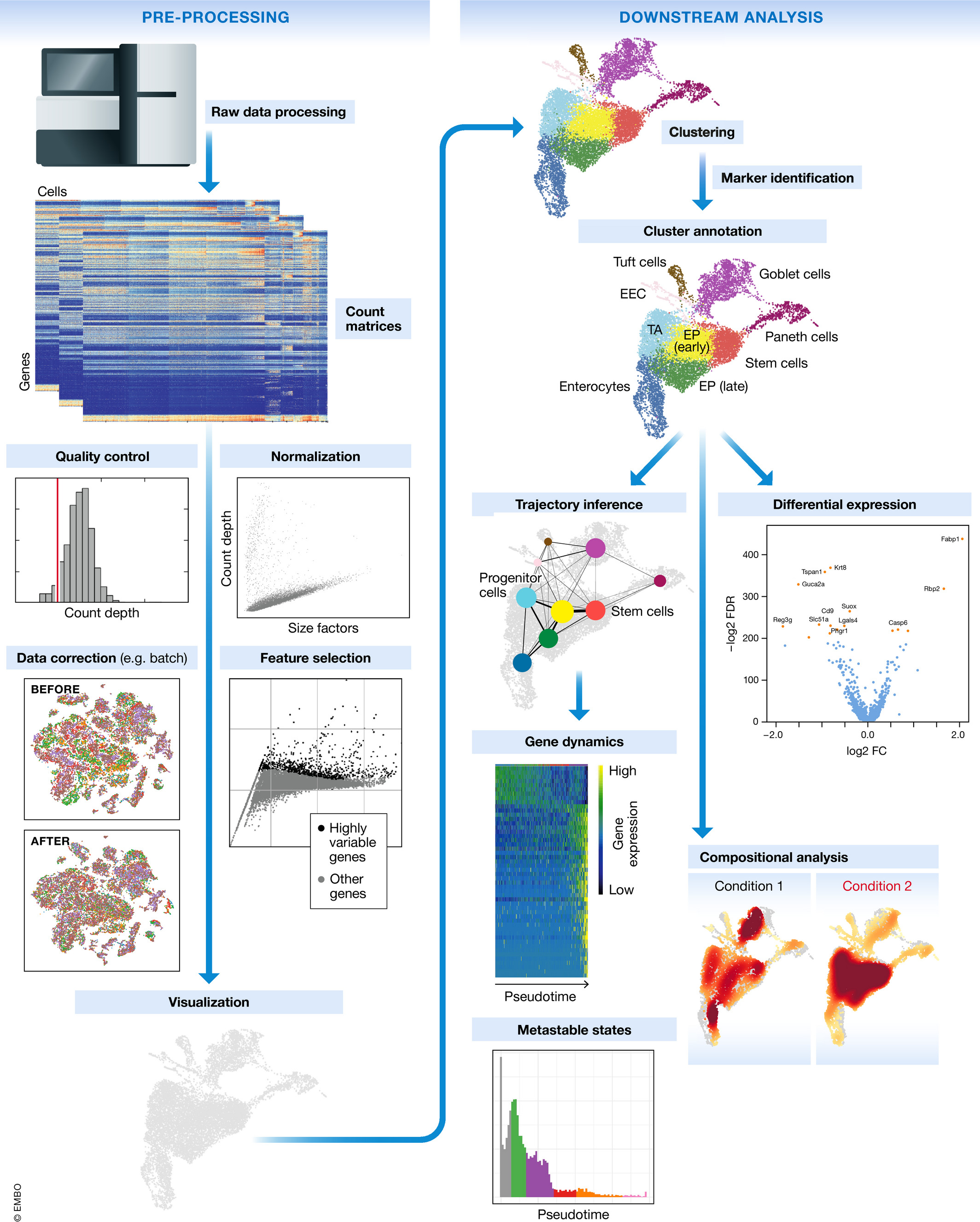}
    \caption{Diagram of a typical single-cell RNA-seq analysis pipeline \cite{Luecken2019}}
    \label{fig:workflow}
\end{figure}

scRNA-seq analysis can be divided into two main stages: pre-processing and downstream analysis. Raw data has to pass through QC to assure no doublets and “bad” cells remain. If data comes from different experiments, batch correction has to be performed to account for the so-called batch-effect. Data (especially, count matrix) also have to be scaled to ease clustering and other related processes. Final stage of pre-processing is feature selection. Every gene represents a feature and using all of them in our analysis would slow it down considerably \cite{Luecken2019}.
\\
Once the data has been pre-processed, it is ready for downstream analysis. Initial stage is clustering, marker identification, and cluster annotation. After this, data is ready for trajectory inference, differential expression and compositional analysis. Trajectory inference offers a dynamic, temporal understanding of gene expression. Compositional analysis is essentially about changes in compositions - cell-identity clusters. Differential expression analysis allows us to see the variety in gene expression across the tissue(s). Downstream analysis does not necessarily include every single one of these stages, although research papers usually include more than one \cite{Luecken2019}. Figure \ref{fig:workflow} gives a detailed illustration of a typical workflow.

\subsection*{Analyzed Papers}
Single-cell analysis has evolved into a considerably large field. As of April, 2020, there are more than 1200 articles in the Single Cell Studies Database. Due to limited time and resources at hand, we picked only two papers from this database, taking into account several factors:  
\begin{itemize}
    \item The paper has to involve single-cell RNA sequencing analysis as one of its major tools of research. That is to say, scRNA-seq analysis has to lead to major findings in the context of the paper.
    \item Most, if not all stages of the scRNA-seq analysis pipeline has to be present in the research. Virtually no single research utilizes every single technique of scRNA-seq analysis, so it is more reasonable to choose our two papers to maximize the coverage.
    \item Single cell data have to be available preferably in raw count matrix format. Count matrix, observations (cells) and variables (genes) constitute the backbone of the scRNA-seq data, no analysis can be performed if even one of them is missing. Cell and gene annotation provide more opportunities for analysis, but they are not imperative.
    \item The tools and platforms used in scRNA-seq analysis have to be documented clearly. We should be able to replicate some of the results presented in the paper (e.g. cell clustering). However, large-scale, especially machine learning algorithms tend to have a stochastic element to them, which means that in order to get some sort of similarity between our results, we need to use exactly the same tools. Otherwise we can end up with virtually distinct outcomes.
\end{itemize}
Vastness of the database allowed us some freedom of choice, which means our preferences also had considerable effect in the choice of papers. We ended up with these two:

\begin{enumerate}
    \item \textit{\cite{Sharma2018}:} In this paper, authors use scRNA-seq to probe variants of chemo-resistance in tumor cells. Sequencing was performed using the C1 Single-Cell Auto Prep IFC (Fluidigm) system. Data has been submitted to NCBI in pre-processed and QC-ed form. Annotations are available for cells, but not for genes. Cell annotations include cell color, patient ID, origin, drug status, and clustering labels.
    \item \textit{\cite{Johnson2020}:} In this paper, authors use scRNA-seq to classify cells found in homeostatic (where there is no need for regeneration) and regenerating mouse digit tips. Sequencing was performed on 10x Genomics Chromium platform. The data has been submitted to NCBI in raw form. We are given 5 different datasets for each stage (each tissue). No annotation is available for genes. However, cell types have been added as supplementary data.
\end{enumerate}

\section*{Results}
\subsection*{\cite{Sharma2018}}
\begin{figure}
     \centering
     \begin{subfigure}[b]{0.48\textwidth}
         \centering
         \includegraphics[width=\textwidth]{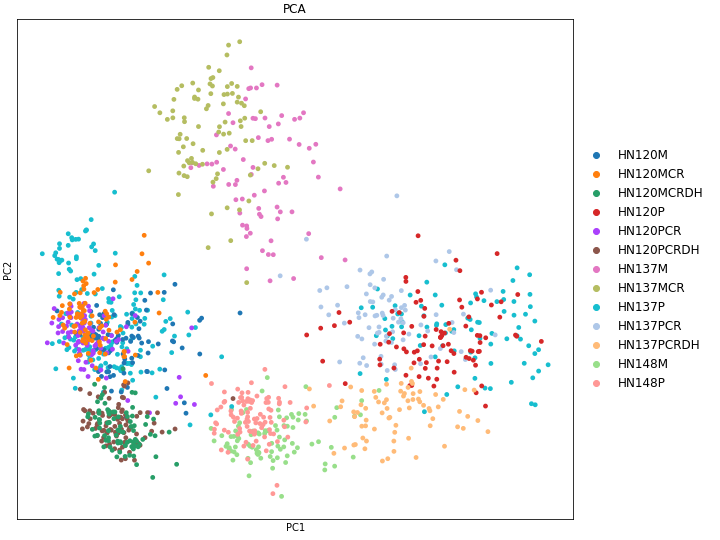}
         \caption{PCA}
         \label{fig:dim_red_1}
     \end{subfigure}
     \hfill
     \begin{subfigure}[b]{0.48\textwidth}
         \centering
         \includegraphics[width=\textwidth]{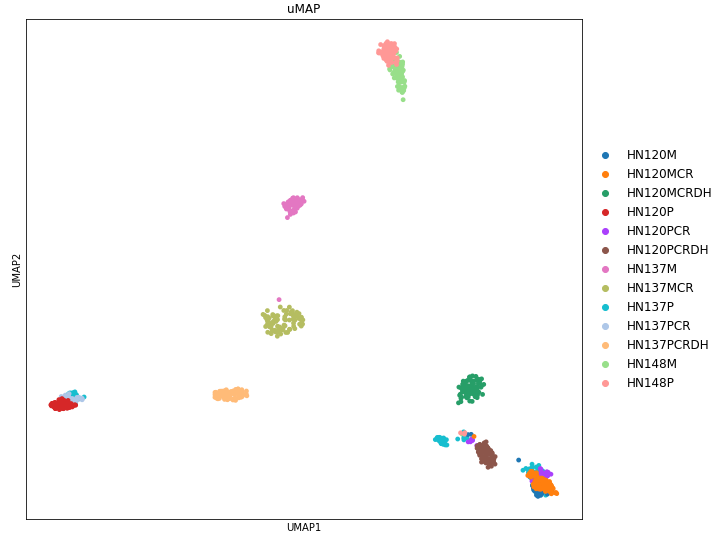}
         \caption{UMAP}
         \label{fig:dim_red_2}
     \end{subfigure}
     \vfill
     \begin{subfigure}[b]{0.48\textwidth}
         \centering
         \includegraphics[width=\textwidth]{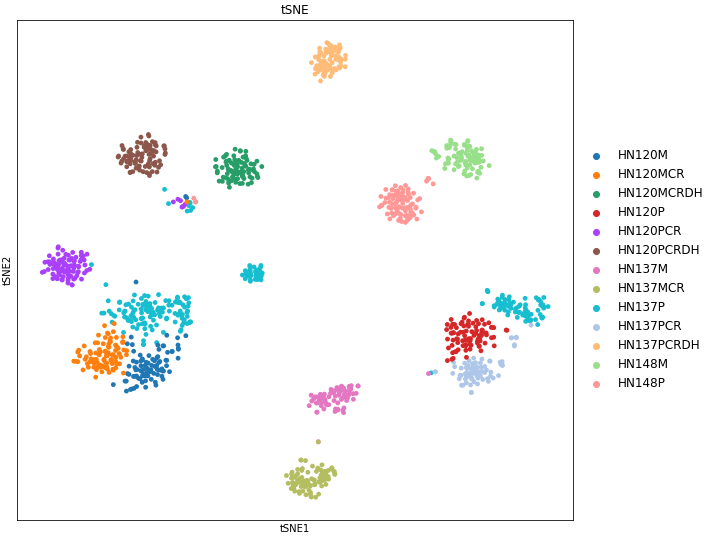}
         \caption{tSNE}
         \label{fig:dim_red_3}
     \end{subfigure}
        \caption{Comparison of dimensionality-reduction algorithms. (Coloring represents cell groups.)}
        \label{fig:dim_red}
\end{figure}

In this article, we are provided with pre-processed count matrix and cell annotations. Therefore, pre-processing algorithms were not tested. Instead, we concentrated on dimensionality-reduction and clustering algorithms. Various clustering metrics were used to compare results of the clustering algorithms. Three different algorithms were probed for dimensionality-reduction:
\begin{enumerate}
    \item Principal Component Analysis (PCA)
    \item Uniform Manifold Approximation and Projection (UMAP)
    \item t-distributed Stochastic Neighbor Embedding (tSNE)
\end{enumerate}
It was discovered that UMAP and tSNE create more distinct clusters, which is useful for visualization purposes, while PCA clusters were more diffuse, hence harder to visually keep track of. We were unable to test whether this apparent advantage of UMAP and tSNE comes with more information loss as there is currently no method to measure information loss for UMAP and tSNE (figure \ref{fig:dim_red}).

\begin{figure}
    \centering
    \includegraphics[width=\textwidth]{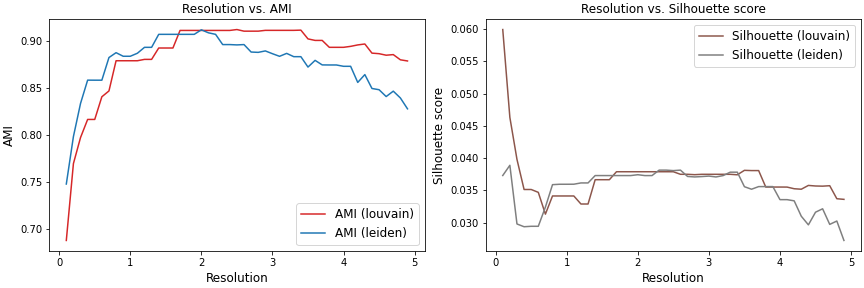}
    \caption{Finding the appropriate resolution for graph-clustering algorithms.}
    \label{fig:res_vs_ami}
\end{figure}

Two different clustering methods were used and their results compared. The first method was k-means (centroid-based) clustering. Principal component analysis was performed on the data, after which scikit-learn implementation of k-means algorithm \cite{pedregosa2018scikitlearn} was applied to it. The second method was graph clustering. We used the Leiden and Louvain algorithms to perform the clustering \cite{Traag2019}. We performed comparisons between these two methods (centroid-based vs. graph clustering), between different algorithms of the same method (Leiden vs. Louvain), as well as between different setups of the same algorithm (k-means clusterings with vs. without PCA).
\\
We initially performed two clusterings using the scikit-learn implementation of k-means algorithm $(K=5)$. One of the clusterings was performed on principal components, while the second clustering was performed on the original data. The results did not differ significantly (See tables \ref{tab:table_1} and \ref{tab:table_2}). We then performed the same comparison among the next two clustering, where we assigned $K = 25$. Increasing the number of clusters widened the gap between PC and non-PC clusterings.
\\
After this, we compared results of Leiden and Louvain algorithms. The Leiden algorithm has been designed based on the Louvain algorithm. It offers faster and more accurate performance \cite{Traag2019}. The difference between the obtained results was not significant enough to reject the null hypothesis (See table \ref{tab:table_3}). (This single case cannot be interpreted as an exhaustive evaluation of Leiden and Louvain algorithms.)
\\
We had to perform preliminary analysis to choose the resolution value for Leiden and Louvain algorithms. Comparing Silhouette score \cite{Rousseeuw1987} with resolution was not useful, as plotting the Silhouette score does not result in an elbow point. Instead, we choose Adjusted Mutual Information score \cite{Vinh2009}, which proved more useful in determining necessary resolution (figure \ref{fig:res_vs_ami}).

Lastly, we compared results of centroid-based and graph clustering methods. We selected K to be 25 for the k-means algorithm. We preferred the Leiden algorithm over Louvain, as the former is known to perform better \cite{Traag2019}. Both clusterings were performed on the principal components.

\begin{table}[]
    \centering
    \begin{tabular}{l|l|l}
        \textbf{Metric} & \textbf{Range} & \textbf{Result}  \\
        \hline
        Adjusted Rand Index & [0,1] & 0.9959 \\
        Jaccard Index & [0,1] & 0.6237 \\
        Adjusted Mutual Information & [0,1] & 0.9933 \\
        Silhouette Coefficient (PCA) & [-1,1] & 0.0588 \\
        Silhouette Coefficient & [-1,1] & 0.0590 \\
        Calinski-Harabasz Index (PCA) & [-,-] & 63.4060 \\
        Calinski-Harabasz Index & [-,-] & 63.4053 \\
    \end{tabular}
    \caption{Comparison of two different k-means clustering results $(K = 5)$. The first clustering has been performed on principal components, while the second clustering has been performed on the original data.}
    \label{tab:table_1}
\end{table}

\begin{table}[]
    \centering
    \begin{tabular}{l|l|l}
        \textbf{Metric} & \textbf{Range} & \textbf{Result}  \\
        \hline
        Adjusted Rand Index & [0,1] & 0.6844 \\
        Jaccard Index & [0,1] & 0.0891 \\
        Adjusted Mutual Information & [0,1] & 0.8281 \\
        Silhouette Coefficient (PCA) & [-1,1] & 0.0273 \\
        Silhouette Coefficient & [-1,1] & 0.0273 \\
        Calinski-Harabasz Index (PCA) & [-,-] & 18.4607 \\
        Calinski-Harabasz Index & [-,-] & 18.3921 \\
    \end{tabular}
    \caption{Comparison of two different k-means clustering results $(K = 25)$. The first clustering has been performed on principal components, while the second clustering has been performed on the original data.}
    \label{tab:table_2}
\end{table}

\begin{table}[]
    \centering
    \begin{tabular}{l|l|l}
        \textbf{Metric} & \textbf{Range} & \textbf{Result}  \\
        \hline
        Adjusted Rand Index & [0,1] & 0.9652 \\
        Jaccard Index & [0,1] & 0.6171 \\
        Adjusted Mutual Information & [0,1] & 0.9688 \\
        Silhouette Coefficient (PCA) & [-1,1] & 0.0359 \\
        Silhouette Coefficient & [-1,1] & 0.0341 \\
        Calinski-Harabasz Index (PCA) & [-,-] & 30.0971 \\
        Calinski-Harabasz Index & [-,-] & 31.2248 \\
    \end{tabular}
    \caption{Comparison of Leiden and Louvain clustering results.}
    \label{tab:table_3}
\end{table}

\begin{table}[]
    \centering
    \begin{tabular}{l|l|l}
        \textbf{Metric} & \textbf{Range} & \textbf{Result}  \\
        \hline
        Adjusted Rand Index & [0,1] & 0.6904 \\
        Jaccard Index & [0,1] & 0.0381 \\
        Adjusted Mutual Information & [0,1] & 0.8617 \\
        Silhouette Coefficient (PCA) & [-1,1] & 0.0359 \\
        Silhouette Coefficient & [-1,1] & 0.0260 \\
        Calinski-Harabasz Index (PCA) & [-,-] & 30.0971 \\
        Calinski-Harabasz Index & [-,-] & 18.4607 \\
    \end{tabular}
    \caption{Comparison of k-means and leiden clustering results. ($K = 15$; both clusterings are performed on principal components.)}
    \label{tab:table_4}
\end{table}

\subsection*{\cite{Johnson2020}}
\begin{figure}
     \centering
     \begin{subfigure}[b]{\textwidth}
         \centering
         \includegraphics[width=0.75\textwidth]{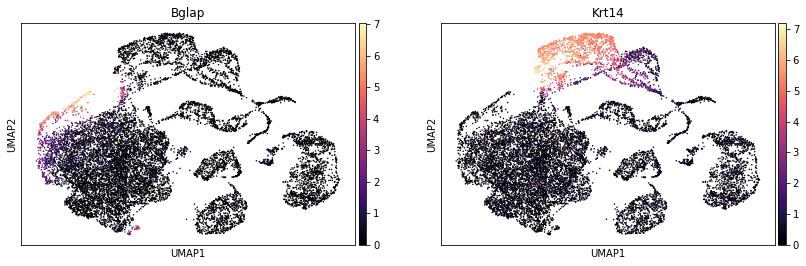}
         \includegraphics[width=0.75\textwidth]{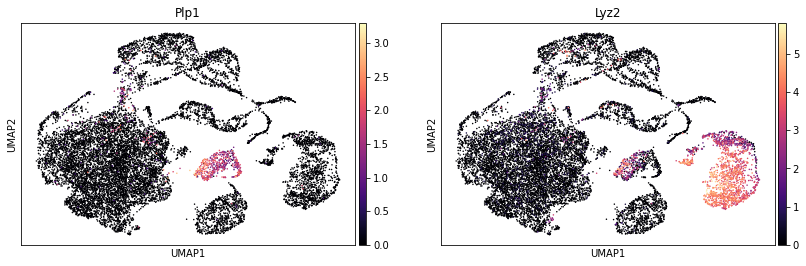}
         \includegraphics[width=0.75\textwidth]{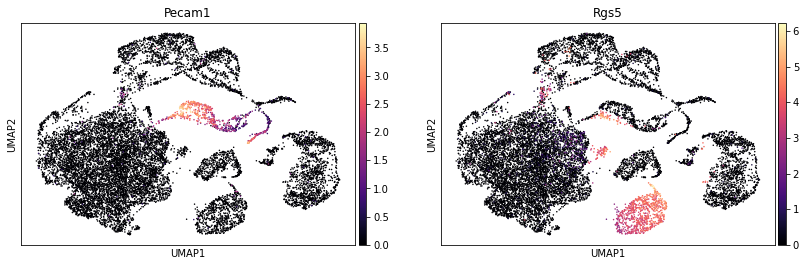}
         \includegraphics[width=0.75\textwidth]{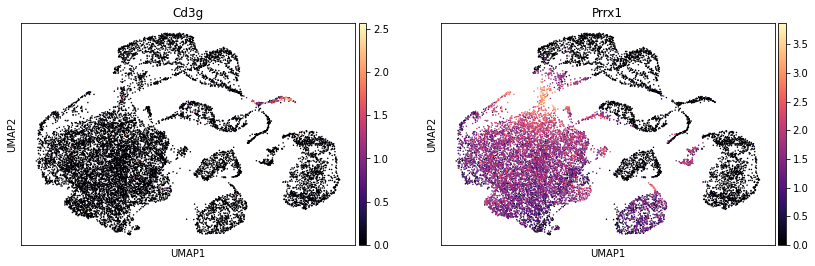}
         \caption{Leiden clusters visualized with help of the UMAP algorithm.}
         \label{fig:scatter_1}
     \end{subfigure}
     \vfill
     \begin{subfigure}[b]{\textwidth}
         \centering
         \includegraphics[width=0.75\textwidth]{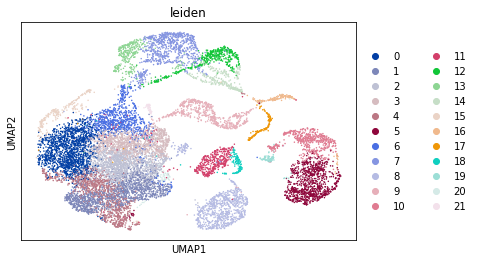}
         \caption{Marker genes, each displayed in individual scatterplot. Scale: expression level (log-normalized UMI).}
         \label{fig:scatter_2}
     \end{subfigure}
        \caption{Cell types are identified by matching clusters \textbf{(b)} with marker genes \textbf{(a)}. In the case of UMAP and scatter plots, it is exceptionally hard. }
        \label{fig:scatter}
\end{figure}

\paragraph{}
In this paper, we are provided with raw count matrix. Single-cell sequencing has been performed on the 10x Genomics Chromium system (Single Cell 3’ v2 and v3 chemistry), which means its coverage is limited. We compared three data visualization methods for marker gene and cell type identification:
\begin{enumerate}
    \item Dimensionality-reduction and scatter plot
    \item Violin plot
    \item Heatmap
\end{enumerate}
After performing Leiden and k-means clustering, the resulting clusters were displayed against expression levels of marker genes. Among the three visualization methods, dimensionality-reduction and scatter plot proved the least effective. Clusters and expression levels of marker genes had to be plotted separately, and then compared. This required too much visual examination on the part of the researcher, and relied on sharp sensory memory, therefore we consider it more prone to human error (figure \ref{fig:scatter}).

\begin{figure}
    \centering
    \includegraphics[width=\textwidth]{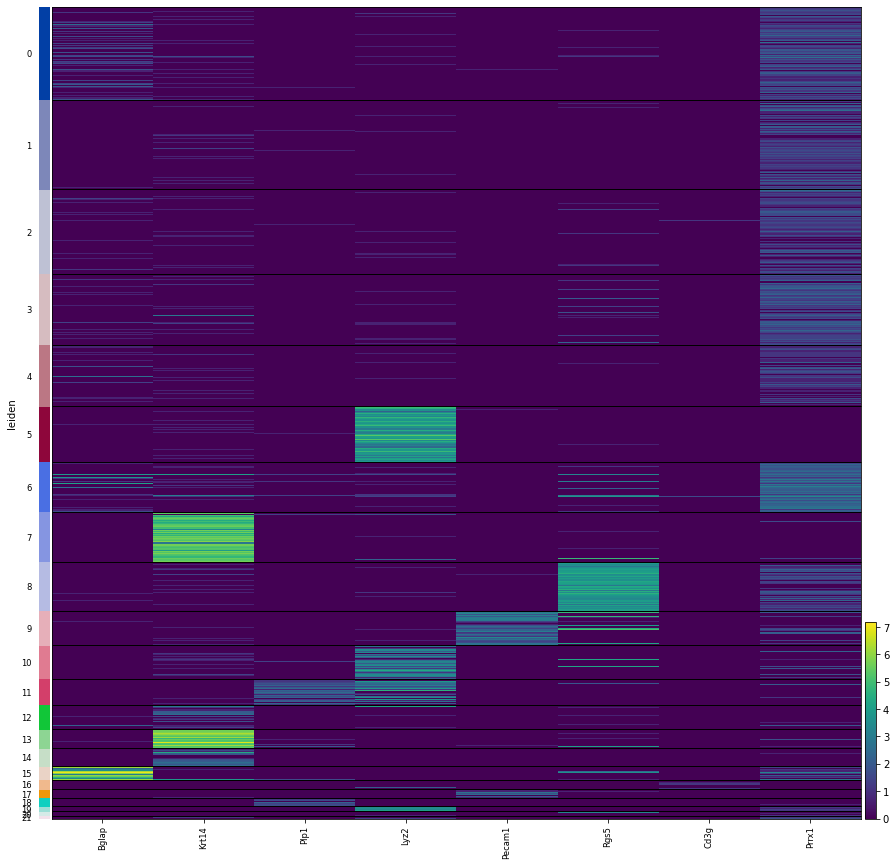}
    \caption{In the case of heatmap, matching clusters and marker genes is relatively easier. However, assigning each cluster to a single marker gene based on the expression levels still causes problems.}
    \label{fig:heatmap}
\end{figure}

Heatmap improves the situation, as it lines up each cluster against expression levels of a marker gene. But expression levels are indicated by a color scale, which makes comparison harder than it should be (figure \ref{fig:heatmap}). We made several mistakes with this method, which we later discovered with help of the violin plot. Violin plot also helped us to make decisions in otherwise undecidable cases.

\begin{figure}
    \centering
    \includegraphics[width=\textwidth]{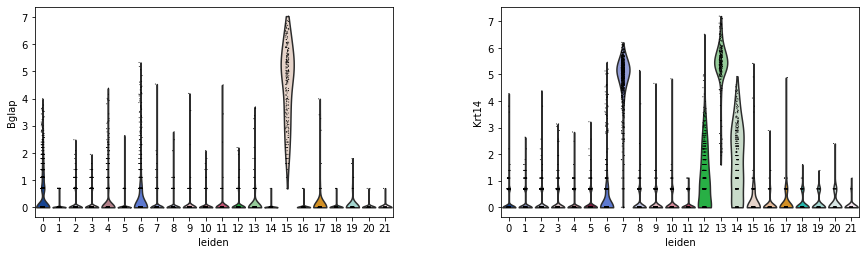}
    \includegraphics[width=\textwidth]{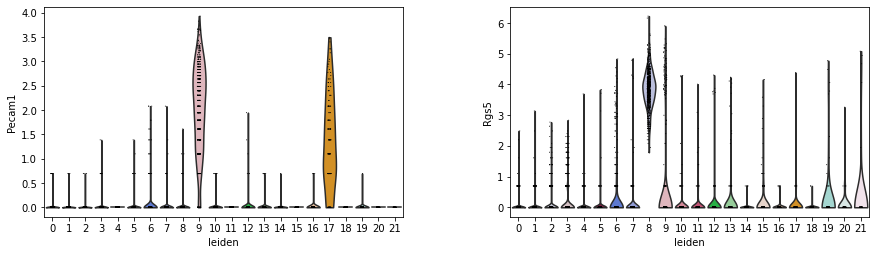}
    \includegraphics[width=\textwidth]{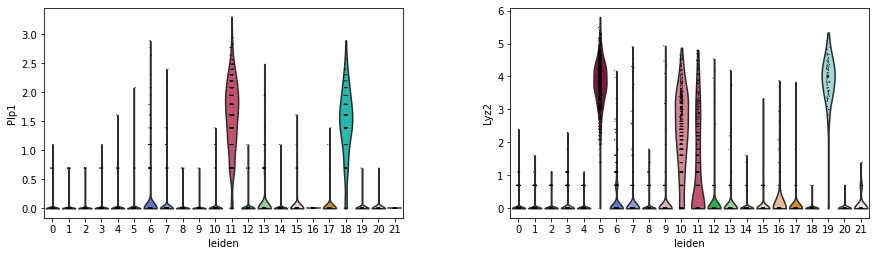}
    \includegraphics[width=\textwidth]{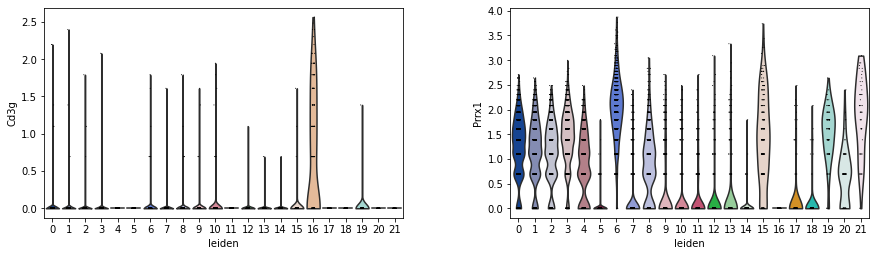}
    \caption{Violin plot decreases the workload significantly, and causes less errors.}
    \label{fig:violin}
\end{figure}

Each violin plot displays clusters against expression levels of a single marker gene. Expression level is indicated by the y axis rather than a color scale. This makes it easier to compare expression levels of a marker gene among different clusters (figure \ref{fig:violin}). Therefore, we consider violin plot the best option to identify cell types with help of marker genes. Final results of cell identification can be found in Supplementary Figures \ref{fig:supp_1}-\ref{fig:supp_3}.

Three algorithms for ranking genes for characterizing groups were also compared. These algorithms are based on three different null-hypothesis tests:
\begin{enumerate}
    \item \textit{Logistic regression} measures potential influence of a factor on the outcome. It allows us to find factors that have the strongest association with the outcome \cite{Tolles2016}. \item \textit{Paired two-sample t-test} is only one of multiple t-tests. It is useful for comparison of paired samples, as it determines whether the mean difference between two sets of observations is significantly different from zero.
    \item \textit{Wilcoxon rank-sum (Mann Whitney U) test} is another method used to compare two samples. It determines whether the probability of X being greater than Y is equal to the probability of Y being greater than X (X and Y being randomly selected values from two different populations).
\end{enumerate}
Initially, we ranked genes for characterizing groups (here, cell types) with help of these three algorithms, and recorded the results separately. Then, we calculated the intersection of the top 100 genes for each group (cell type) computed by:
\begin{enumerate}
    \item Wilcoxon rank-sum test vs. paired two-sample t-test
    \item Wilcoxon rank-sum test vs. logistic regression
    \item Paired two-sample t-test vs. logistic regression
\end{enumerate}
We found out that Wilcoxon signed-rank test and paired two-sample t-test were closer to each other than to logistic regression.

\begin{figure}
     \centering
     \begin{subfigure}[b]{\textwidth}
         \centering
         \includegraphics[width=0.7\textwidth]{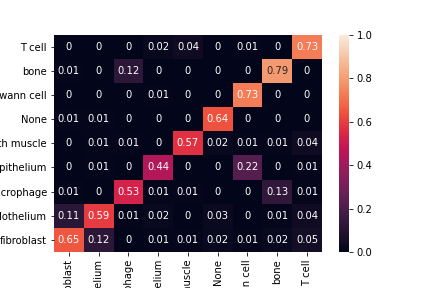}
         \caption{Paired two-sample t-test vs. logistic regression.}
         \label{fig:wilcoxon_1}
     \end{subfigure}
     \vfill
     \begin{subfigure}[b]{\textwidth}
         \centering
         \includegraphics[width=0.7\textwidth]{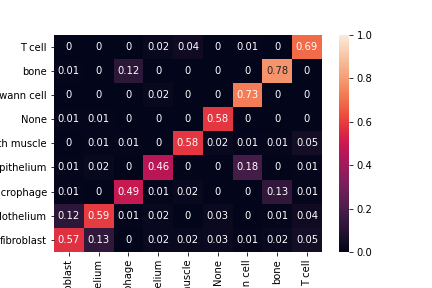}
         \caption{Wilcoxon rank-sum test vs. logistic regression.}
         \label{fig:wilcoxon_2}
     \end{subfigure}
     \vfill
     \begin{subfigure}[b]{\textwidth}
         \centering
         \includegraphics[width=0.7\textwidth]{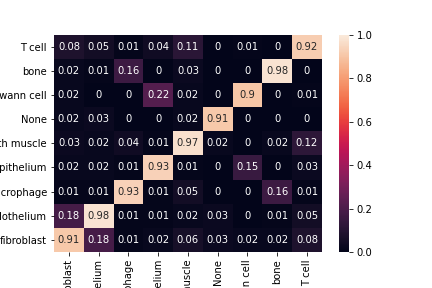}
         \caption{Wilcoxon signed-rank test vs. paired two-sample t-test.}
         \label{fig:wilcoxon_3}
     \end{subfigure}
     \caption{Intersection of top 100 genes (ranked for characterizing cell types), computed by (a) Wilcoxon rank-sum test, (b) paired two-sample t-test, and (c) logistic regression. (Scale: intersection size)}
     \label{fig:wilcoxon}
\end{figure}

\section*{Conclusion and outlook}
We compared different algorithms used in several stages of single-cell RNA sequencing analysis. Main concentration was on dimensionality-reduction, clustering, visualization, and ranking genes for characterizing groups. In some cases, we were able to make clear recommendations regarding these algorithms. However, in most cases, we managed to perform only surface-level analysis, with no basis for a recommendation.
\\
When performing dimensionality reduction for visualization purposes, UMAP or tSNE are suggested over PCA, as they result in more distinct clusters. The higher cluster count is selected in k-means clustering, the more significant effect PCA has on the result. No significant difference was found in performance of Leiden and Louvain algorithms. When comparing expression levels of marker genes and results of clustering, violin plot is suggest over both dimensionality-reduction/scatter plot and heatmap.
\\
Lack of a method to measure information loss for UMAP and tSNE is serious challenge that hindered our research process. Solution of this problem would help us decide between PCA, UMAP, and tSNE algorithms for dimensionality-reduction.

\section*{Methods and tools}
Our entire code was written in Python (v3..8.5). Scanpy (v1.7.1), scikit-learn (v0.24.1), leidenalg (v0.8.3), numpy (v1.20.2), and pandas (v1.2.3) packages have been used extensively. Biological data has been converted into an AnnData (v0.7.5) object before being processed. Other packages involved in the project are: seaborn (v0.11.1), matplotlib (v.3.4.1), and scipy (v1.6.3).
\\
The following functions have been used in our analysis:
\begin{enumerate}
\item scanpy.pp.highly\_variable\_genes function has been used to find highly variable genes (n\_top\_genes = 3000). 
\item scanpy.pp.pca (n\_comps = 75 in the first paper, n\_comps=16 in the second paper) and sklearn.decomposition.PCA have been used to perform PCA. 
\item scanpy.pp.neighbors (n\_pcs = 75 in the first paper, default in the second paper) has been used to compute neighborhood graph of observations. 
\item scanpy.pl.umap, and scanpy.pl.tsne have been used for dimensionality reduction. 
\item sklearn.metrics.adjusted\_rand\_score has been used to compute adjusted rand index. 
\item sklearn.metrics.jaccard\_score has been used to calculate Jaccard index. \item sklearn.metrics.silhouette\_score has been used to calculate silhouette score.
\item sklearn.metrics.adjusted\_mutual\_info\_score has been used to calculate adjusted mutual information.
\item sklearn.metrics.calinski\_harabasz\_score has been used to calculate Calinski-Harabasz index. 
\item sklearn.cluster.KMeans (n\_clusters = 5 and 25 in the first paper, n\_clusters = 22, 17, 15, 19) has been used for k-means clustering. 
\item sc.tl.louvain has been used to perform Louvain clustering. sc.tl.leiden (resolution = 1 in the first paper, resolution = 0.45 and 0.6 in the second paper) has been used to perform Leiden clustering.
\end{enumerate}

Code and accompanying notes have been written in Jupyter Notebooks. They are available on GitHub: \href{https://github.com/ceferisbarov/scRNA-seq}{https://github.com/ceferisbarov/scRNA-seq}

\begin{table}[h]
    \centering
    \resizebox{\textwidth}{!}{
    \begin{tabularx}{\columnwidth}{X|X|X}
        RESOURCE & SOURCE & DOCUMENTATION \\
        \hline
        Python (Version 3.8.5) & Python Software Foundation & \href{https://docs.python.org/}{docs.python.org}\\
        Scanpy (Version 1.7.1) & \cite{Wolf2018} & \href{https://scanpy.readthedocs.io/en/stable/api.html}{scanpy.readthedocs
        .io/en/stable/api.html} \\
        Anndata (Version 0.7.5) & \cite{Wolf2018} & \href{https://anndata.readthedocs.io/en/latest/api.html}{anndata.readthedocs
        .io/en/latest/api.html} \\
        leidenalg (Version 0.8.3) & \cite{Traag2019} & \href{https://leidenalg.readthedocs.io/en/stable/intro.html}{leidenalg.readthedocs
        .io/en/stable/intro.html} \\
        scikit-learn (Version 0.24.1) & \cite{pedregosa2018scikitlearn} & \href{https://scikit-learn.org/stable/modules/classes.html}{scikit-learn.org/stable/
        modules
        /classes.html} \\
        numpy (Version 1.20.2) & \cite{Harris2020} & \href{https://numpy.org/doc/stable/reference/}{numpy.org/doc/stable
        /reference}
    \end{tabularx}}
    \label{tab:my_label}
\end{table}

\bibliographystyle{apalike} 
\bibliography{bibliography}
    
\newpage
\section*{Supplementary information}
\renewcommand{\thefigure}{S\arabic{figure}}
\setcounter{figure}{0}
\begin{figure}[h]
    \centering
    \includegraphics[width=\textwidth]{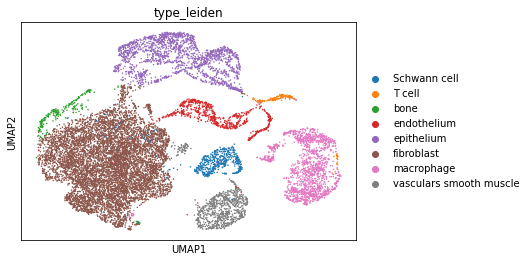}
    \caption{Visualization of predicted cell types with help of a scatter plot (UA). Cells were clustered with help of the Leiden algorithm and each cluster was assigned a cell type based on expression levels of marker genes.}
    \label{fig:supp_1}
\end{figure}

\begin{figure}
    \centering
    \includegraphics[width=\textwidth]{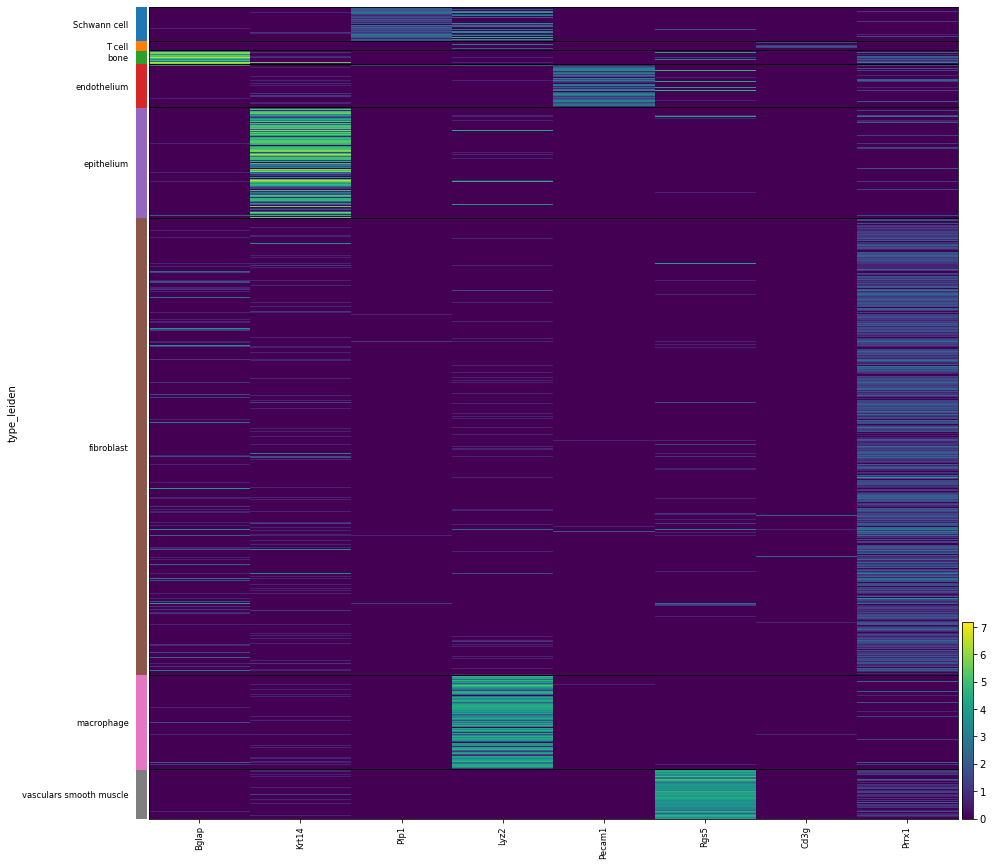}
    \caption{Visualization of predicted cell types with help of a heatmap (UA).}
    \label{fig:supp_2}
\end{figure}

\begin{figure}
    \centering
    \includegraphics[width=\textwidth]{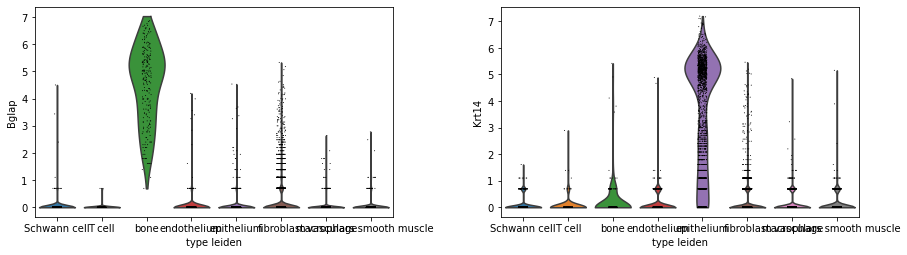}
    \includegraphics[width=\textwidth]{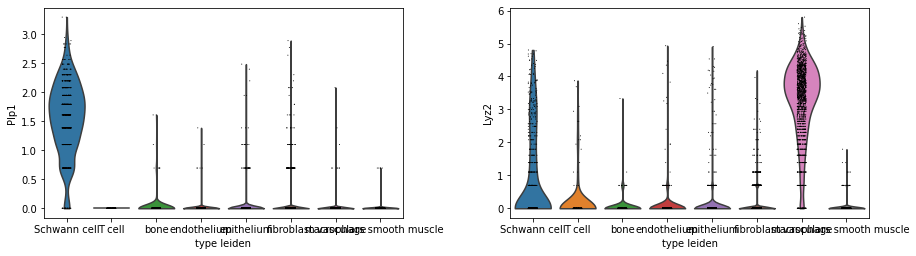}
    \includegraphics[width=\textwidth]{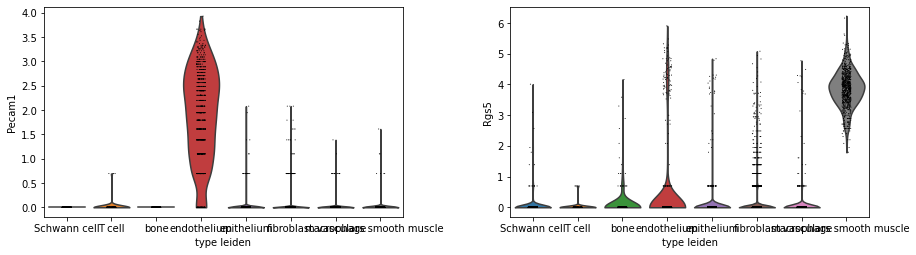}
    \includegraphics[width=\textwidth]{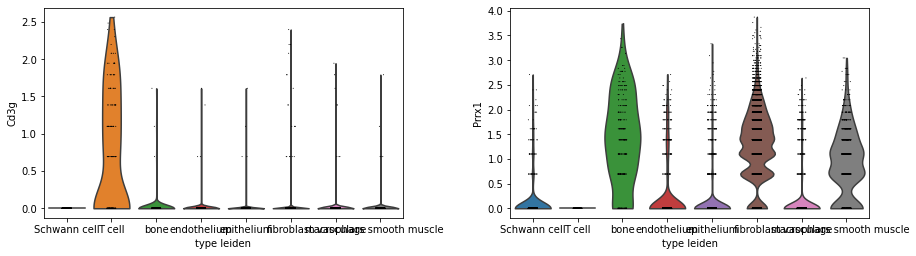}
    \caption{Visualization of predicted cell types with help of a violin plot (UA).}
    \label{fig:supp_3}
\end{figure}
\end{document}